\documentclass[12pt]{article}
\usepackage{amsmath,epsf,amssymb,latexsym,cite}
\usepackage[ps,dvips,matrix,arrow,frame,import,curve,color]{xy}

\newif\iffigs\figstrue

\DeclareFontFamily{U}{rsf}{}
\DeclareFontShape{U}{rsf}{m}{n}{
  <5> <6> rsfs5 <7> <8> <9> rsfs7 <10-> rsfs10}{}
\DeclareMathAlphabet\Scr{U}{rsf}{m}{n}

\def\pplogo{\vbox{\kern-\headheight\kern -29pt
\halign{##&##\hfil\cr&{%\sc
\ppnumber}\cr\rule{0pt}{2.5ex}&\ppdate\cr}
}}
\makeatletter
\def\ps@firstpage{\ps@empty \def\@oddhead{\hss\pplogo}%
  \let\@evenhead\@oddhead % in case an article starts on a left-hand page
}
\def\maketitle{\par
 \begingroup
 \def\thefootnote{\fnsymbol{footnote}}
 \def\@makefnmark{\hbox{$^{\@thefnmark}$\hss}}
 \if@twocolumn
 \twocolumn[\@maketitle]
 \else \newpage
 \global\@topnum\z@ \@maketitle \fi\thispagestyle{firstpage}\@thanks
 \endgroup
 \setcounter{footnote}{0}
 \let\maketitle\relax
 \let\@maketitle\relax
 \gdef\@thanks{}\gdef\@author{}\gdef\@title{}\let\thanks\relax}
\makeatother

\def\eh{\hat{\epsilon}}

\begin{document}
\setcounter{page}0
\def\ppnumber{\vbox{\baselineskip14pt
\hbox{SU-ITP-05/23}
\hbox{hep-th/0506132}}}
\def\ppdate{June 2005} \date{}

\title{\LARGE Solving the eigenvalue problem arising from the adjoint sector of the $c=1$ matrix model\\[10mm]}
\author{
Lukasz Fidkowski\\[2mm]
\normalsize Departments of Physics\\
\normalsize Stanford University
}

{\hfuzz=10cm\maketitle}

\def\Large{\large}
\def\LARGE{\large\bf}

\vskip 1cm

\begin{abstract}
We solve the non-local eigenvalue problem that arose from consideration of the adjoint sector of the $c=1$ matrix model in \cite{Maldacena:2005hi}.  We obtain the exact wavefunction and a scattering phase that matches the string theory calculation.
\end{abstract}

\vfil\break

%%%%%%%%%%%%%%%%%%%%%%%%%%%%%%%%%%%%%%%%%%%%%%%%%%%%%%%%%%%%%%%%

\section{Introduction}    \label{s:intro}

The motivation for this paper is to solve exactly an eigenvalue problem that arises in \cite{Maldacena:2005hi} and show that the exact scattering phase we obtain matches the string theory calculation.  This provides further confirmation of the duality between the nonsinglet sectors of the $SU(N)$ matrix model and time dependent long string configurations in $c=1$ string theory.

The problem is 

\begin{equation}
- \int_{-\infty}^{\infty} d\tau' {h\left( \tau' \right) \over 4 \sinh^2 {\tau - \tau' \over 2}} - {\tau \over \tanh \tau} h(\tau) = \pi \eh h(\tau) \label{eq:first}
\end{equation}which can also be written as

\begin{equation}
\pi \eh h(\tau) = \left( {\pi k \over \tanh \pi k} - {\tau \over \tanh \tau} \right) h(\tau)
\end{equation}where $k=-i \partial_\tau$.  It arises in general when one tries to compute the meson (i.e. adjoint sector) spectrum of the $SU(N)$ matrix model \cite{MarOno}.  As opposed to the singlet sector, which for a general potential reduces to $N$ independent fermions, the non-singlet sector hamiltonian includes interactions between the different eigenvalues.  This interaction piece can be treated perturbatively in $1/N$ \cite{MarOno}, and when, as in \cite{Maldacena:2005hi}, one is interested in the double scaling limit around an inverted harmonic oscillator potential, the problem reduces to (\ref{eq:first}).

We obtain the exact eigenstate of the above problem for any $\eh$ by summing up a perturbation series.  The answer we get looks very simple, and there may be another, easier way of obtaining it.

\section{Summing the Perturbation Series}

\subsection{Set-up}

We set up the problem following the notation of \cite{Maldacena:2005hi}.  The eigenvalue equation (\ref{eq:first}) can be written in Fourier space as:

\begin{equation}
\left( {\pi k \over \tanh \pi k} - {i \partial_k \over \tanh i \partial_k} \right) h(k) = \pi \eh h(k).\label{eq:eigen}
\end{equation}To start we take $\eh \ll 0$.  In $\tau$ space this corresponds to a wave with very negative energy bouncing off an essentially linear potential.  In $k$ space the picture is a wave with very positive energy which barrels through the potential very fast, so that its kinetic term is essentially linear, i.e. having a relativistic dispersion:

\begin{equation}
\eh h_0 (k) = \left( {k \over \tanh \pi k} - {i \over \pi} \partial_k \right) h_0 (k).
\end{equation}The solution is

\begin{equation}
h_0(k) = \exp \left( i \pi \hat \epsilon k -  i \pi \int^k dk' { k' \over \tanh \pi k' } \right) \label{eq:pert}
\end{equation}

\begin{equation}
h_0(k) ~ \longrightarrow \exp\left( i \pi \hat \epsilon k \mp {i \pi \over 2} k^2 \mp i \pi/12
\right) ~~~~~~{\rm as} ~~~~~~k\to \pm \infty
\end{equation}For large $\tau$ we can Fourier transform back and obtain (up to overall phase) \footnote{In \cite{Maldacena:2005hi} there appears to be a typo in the corresponding formula (B.7).  Specifically, in (B.6) there is a typo omitting the $\mp$ in front of the $i \pi / 12$, and also it is not taken into account that because of the $\mp$ in front of the term quadratic in $k$ in (B.6) the determinants arising from the gaussian integrals have a relative factor of $i$.}

\begin{equation}
h_0(\tau) \sim
e^{ - i { 1 \over 2 \pi }(\tau + \pi \hat \epsilon)^2 - {i \pi}/6 } -
 e^{ i { 1 \over 2 \pi} (\tau + \pi \hat \epsilon)^2 + {i \pi}/6 } ~,~~~~~~~\tau \gg 1.
\end{equation}Moving away from the strict limit $\eh \ll 0$ we will have a scattering phase

\begin{equation}
h_0(\tau) \sim
e^{i \delta / 2} e^{ - i { 1 \over 2 \pi }(\tau + \pi \hat \epsilon)^2 - {i \pi}/6} - e^{-i \delta / 2}
 e^{ i { 1 \over 2 \pi} (\tau + \pi \hat \epsilon)^2 + {i \pi}/6} ~,~~~~~~~\tau \gg 1
\end{equation}The idea is now to compute the phase $\delta$ by expanding the wave-function in powers of $e^{2\pi\eh}$.  In \cite{Maldacena:2005hi} the first correction is computed and at that order matches the conjectured exact result

\begin{equation}
\delta(\hat \epsilon) = - \int^{\hat \epsilon}_{-\infty} \left(
  { \pi \epsilon' \over \tanh \pi \epsilon'} + \pi \epsilon' \right). \label{eq:phase}
\end{equation}The perturbed wave-function that is obtained there has a singularity at $k=0$.  This presumably reflects the fact that in $\tau$ space the wave-function is unstable to tunneling under the barrier.  When we sum up all the corrections the singularities will turn out to add up to a branch cut, and when we anti-symmetrize properly the branch cut will disappear.  We will then explicitly verify the solution.

For now, we write the purported solution:
\begin{equation}
h(k) = \exp \left( i \pi \hat \epsilon k -  i \pi \int_0^k dk' { k' \over \tanh \pi k' } + e^{2\pi\eh} v_1 + e^{4\pi\eh} v_2 + \cdots \right)\label{eq:purport}
\end{equation}and plug it in.  We could have a priori made the expansion a prefactor multiplying the exponential, but a little experimentation shows that it is easiest to do the expansion inside the exponent.  As in \cite{Maldacena:2005hi} we obtain for the first correction:

\begin{equation}
{1 \over i \pi} \partial_k v_1 + {2 \over (e^{\pi k} - e^{- \pi k})^2} \left(\eh - {k-2i \over \tanh \pi k}\right) = 0
\end{equation}

\begin{equation}
v_1 = -2\pi i \int^k \left( \eh - {k' - 2i \over \tanh \pi k'} \right) {dk' \over (e^{\pi k'} - e^{-\pi k'})^2}
\end{equation}Continuing and computing $v_2$ leads us to conjecture that the general correction will take the form

\begin{equation}
v_n = -2 \pi i \int^k \left( a_n \eh - {a_n k' - 2 b_n i \over \tanh \pi k'} \right) {dk' \over (e^{\pi k'} - e^{-\pi k'})^{2n}}
\end{equation}where $a_n$ and $b_n$ are unknown coefficients.  Plugging into (\ref{eq:eigen}) we obtain recursion relations that determine all the $a_n$ and $b_n$.  The goal in the next subsection is to solve these.

\subsection{Solving the Recursion Relations}

First, it will be useful to note that

\begin{equation}
\begin{split}
v_n(k-2i) - v_n(k) =2\pi i a_n \int_k^{k-2i} {k' \over \tanh \pi k'} {dk' \over (e^{\pi k'} - e^{-\pi k'})^{2n}}\\
=\left[ 2\pi i a_n k' \left( -{1 \over 2 \pi n} \right) {1 \over (e^{\pi k'} - e^{-\pi k'})^{2n}} \right]_k^{k-2i}\\
=\left(-{2 a_n \over n} \right) {1 \over (e^{\pi k} - e^{-\pi k})^{2n}}.
\end{split}
\end{equation}and thus

\begin{equation}
v_n(k-2mi) - v_n(k) = \left(-{2 m a_n \over n} \right) {1 \over (e^{\pi k} - e^{-\pi k})^{2n}} \label{eq:diff}
\end{equation}To keep expressions manageable, we define generating functions that encode the information in the $a_n$ and $b_n$:

\begin{equation}
\begin{split}
F(x) = \sum_{n \geq 1} {a_n \over n} x^n\\
G(x) = \sum_{n \geq 1} b_n x^n.
\end{split}
\end{equation}We see that the combination $e^{2\pi \eh} / (e^{\pi k} - e^{-\pi k})^2$ appears a lot, so henceforth we set

\begin{equation}
x(k) = {e^{2\pi \eh} \over (e^{\pi k} - e^{-\pi k})^2}
\end{equation}We will suppress the argument $k$ when no ambiguity can arise.  To avoid messy equations, let us label by $A(k)$ the derivative with respect to $k$ of the exponent in (\ref{eq:purport}):

\begin{equation}
A(k) = \eh - {k \over \tanh \pi k} - 2 \sum_{n \geq 1} x^n \left( a_n \eh - {a_n k - 2 b_n i \over \tanh \pi k} \right),
\end{equation}For conciseness we have omitted the dependence of $A(k)$ on $\eh$ in the notation.  Now, by expanding the $\tanh i \partial_k$ in (\ref{eq:eigen}) and using (\ref{eq:diff}) we obtain the following

\begin{equation}
\begin{split}
\eh - {k \over \tanh \pi k} = -A(k) + 2 \sum_{m \geq 0} x^m \exp \left({-m \sum_{n\geq1} {2 a_n \over n} x^n }\right) A(k-2mi)\\
= -A(k) + 2 \sum_{m \geq 0} \left( x \exp \left({-\sum_{n\geq1} {2 a_n \over n} x^n} \right) \right)^m \left[ A(k) + {2mi \over \tanh \pi k} + \sum_{n \geq 1} \left( -2x^n\right) {2mi a_n \over \tanh \pi k} \right]\\
= -A(k) + {2 \over 1 - x e^{-2F(x)}} A(k) + {2 x e^{-2F(x)} \over (1 - x e^{-2F(x)})^2} \left({2i \over \tanh \pi k} - {4i \over \tanh \pi k} x F'(x) \right).
\end{split} \label{eq:long}
\end{equation}We also note that

\begin{equation}
A(k) = \left(\eh - {k \over \tanh \pi k}\right) \left(1 - 2x F'(x) \right) - {4i \over \tanh \pi k} G(x)
\end{equation}so that (\ref{eq:long}) becomes

\begin{equation}
\begin{split}
\eh - {k \over \tanh \pi k} = \left(1+xe^{-2F} \over 1-xe^{-2F} \right) \left[\left(\eh - {k \over \tanh \pi k}\right) \left(1-2xF' \right) - {4i \over \tanh \pi k} G \right]\\
+ {4ixe^{-2F} \over \left(1-xe^{-2F}\right)^2} {1 \over \tanh \pi k} \left(1 - 2xF' \right). \label{eq:diffeq}
\end{split}
\end{equation}Let us solve this equation.  First off, the terms proportional to $\eh - {k \over \tanh \pi k}$ must be equal, so that

\begin{equation}
\left(1-2xF'(x)\right) \left(1+xe^{-2F(x)}\right) = 1-xe^{-2F(x)}
\end{equation}which can be reduced to the form

\begin{equation}
2 \left(e^{-2F(x)}\right)^2 + \left(e^{-2F(x)}\right)' + x {e^{-2F(x)}}\left(e^{-2F(x)}\right)' = 0
\end{equation}This is solved by

\begin{equation}
F(x) = -{1 \over 2} \log \left(1+2x-\sqrt{1+4x} \over 2x^2\right)
\end{equation}Setting the remainder of the terms in (\ref{eq:diffeq}) equal to each other we get

\begin{equation}
\begin{split}
G(x) = {x e^{-2F} \left(1-2xF'(x)\right) \over 1-x^2 e^{-4F}} = {x \over 1+4x}.
\end{split}
\end{equation}

\subsection{The Solution}

Now that we've solved for the $a_n$ and $b_n$ we plug in to get the final form of the solution.  Integrating by parts we first obtain

\begin{equation}
\begin{split}
v_n = -2\pi i a_n\left(\eh - {1 \over 2 \pi n}\right) \int^k {dk' \over (e^{\pi k'} - e^{-\pi k'})^{2n}}\\
-{i\over n}\left(a_n k - 2 i b_n\right){1 \over (e^{\pi k} - e^{-\pi k})^{2n}}
\end{split}
\end{equation}Summing,

\begin{equation}
\sum_{n \geq 1} e^{2 \pi n \eh} v_n = -2\pi i \eh \int^k{dk' x(k') F' \left(x(k') \right)} + i \int^k {F \left(x(k')\right)} - {1 \over 2}\log \left(1+4x(k)\right) -
i k F\left(x(k)\right).
\end{equation}After some easy algebraic manipulation of the terms above that involve $F(x)$, we get that the total exponent in (\ref{eq:purport}) is

\begin{equation}
i \pi \eh k - i \pi \int^k {k' dk' \over \tanh \pi k'} + \sum_{n \geq 1} e^{2 \pi n \eh} v_n = \pi i \int^k {dk' \over \sqrt{1+4x}} \left(\eh - {k' \over \tanh \pi k'}\right) - {1 \over 2} \log \left(1+4x\right).
\end{equation}So the purported solution is

\begin{equation}
h(k) = {1\over\sqrt{1+4x}} \exp \left\{ \pi i \int^k {dk' \over \sqrt{1+4x}} \left(\eh - {k' \over \tanh \pi k'}\right) \right\}
\end{equation}Substituting in ${e^{2\pi \eh} / (e^{\pi k} - e^{-\pi k})^2}$ for $x$, we obtain

\begin{equation}
h(k) = {\sinh {\pi k} \over \sqrt{\sinh^2 {\pi k} + e^{2 \pi \eh}}} \exp \left\{ \pi i \int^k {dk' \sinh {\pi k'} \over \sqrt{\sinh^2 {\pi k'} + e^{2 \pi \eh}}} \left(\eh - {k' \over \tanh \pi k'} \right) \right\}.\label{eq:soln}
\end{equation}.

\subsection{Branches}
There are some problems with the above solution.  For one, the square root will have a branch cut, and we will have to decide where it goes.  The most naive choice of the branch cut (which avoids the real $k$ axis altogether) makes the exponent in (\ref{eq:soln}) an even function of $k$, and makes the solution odd in $k$.  This means that for both large positive $k$ and large negative $k$ the wavefunction is outgoing (or incoming, depending on conventions), and that we have two purported independent solutions (for the two signs of the square root).  There is no local conservation so this might not be a problem, though by building wavepackets one might run into problems with unitarity (have two packets that collide and disappear).

It should be noted, however, that this choice of branch cut makes the exponent in the perturbed solution (\ref{eq:soln}) not converge to that of the unperturbed solution (\ref{eq:pert}) as $\eh$ goes to minus infinity.  Instead, from the standpoint of the perturbation expansion it is more natural to choose the branch cut to bisect the real $k$ axis at $k=0$.   Also, when computing the first perturbative correction to the phase shift, \cite{Maldacena:2005hi} integrates around the singularity at $k=0$; as successive more singular terms in the expansion pile up at $k=0$, they fatten up into a branch cut that bisects the real $k$ axis at $k=0$.  With this choice of branch cut, the function one obtains is not smooth at $k=0$, and is not odd in $k$.  However, after anti-symmetrization, it becomes odd and smooth at $k=0$. In fact, it becomes analytic on the entire complex $k$ plane, a fact which we will need in verifying that it indeed satisfies (\ref{eq:eigen}).

Before we prove these things, let us summarize.  We choose the branch cut in the wavefunction to go through $k=0$, in order to have a solution that approaches the unperturbed solution when $\eh$ goes to minus infinity.  After anti-symmetrization we will show that we obtain an analytic wavefunction.  By plugging in, we independently verify in the next section that it does satisfy (\ref{eq:eigen}).  It is the only combination of the two branches that can be verified to satisfy (\ref{eq:eigen}) using the method in the next section.  So even though there are other branches that give different wavefunctions that are smooth for real $k$ (such as the ones discussed in the first paragraph of this section), they cannot be verified using the methods of the next section.  We believe that these other branches in fact do not satisfy (\ref{eq:eigen}), though we do not prove this uniqueness statement.

Let us now proceed systematically.  First, the integrand in the exponent of (\ref{eq:soln}) has branch points whenever $\sinh^2 k = -e^{2 \pi \eh}$ and defines a Riemann surface which is a double cover of the complex $k$-plane.  What is not so obvious, but nevertheless true, is that the whole solution (\ref{eq:soln}) is a well defined function on this Riemann surface.  Let us see why this is so.  First, we need a branch cut structure for $\sqrt{\sinh^2 {\pi k} + e^{2 \pi \eh}}$ - for definiteness, take the one discussed above (the one that goes through $k=0$).  To continue, it will be useful to relate $h(k+2i)$ to $h(k)$.  We have

\begin{equation}
h(k+2i) = \exp \left\{ \pi i \int_k^{k+2i} {dk' \sinh {\pi k'} \over \sqrt{\sinh^2 {\pi k'} + e^{2 \pi \eh}}} \left(\eh - {k' \over \tanh \pi k'} \right) \right\} h(k).
\end{equation}For definiteness take $k$ real and positive.  The integral in the exponent is

\begin{equation}
i \int_k^{k+2i} {\eh d(\cosh \pi k') \over \sqrt{\cosh^2 \pi k' - (1 - e^{2 \pi \eh})}} - i \int_k^{k+2i} {k' d(\sinh \pi k') \over \sqrt{\sinh^2 \pi k' + e^{2 \pi \eh}}}.
\end{equation}The first term is easily evaluated to be $-2 \pi \eh$.  The second one is trickier.  Letting $u = \sinh \pi k'$ and $u_0 = \sinh \pi k$ we see that it is equal to

\begin{equation}
-{i \over \pi} \int {\log \left( u + \sqrt{u^2 + 1} \right) du \over \sqrt{u^2 + e^{2 \pi \eh}}} \label{eq:cuts}
\end{equation}

\begin{figure}[h!]
\centering\leavevmode\epsfysize= 5.5 cm \epsfbox{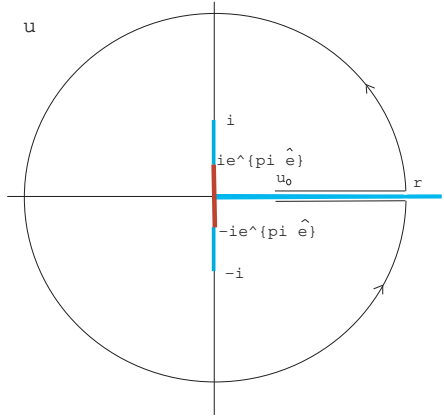}

\

\caption[w1] {The contour of integration and branch cut structure for (\ref{eq:cuts}).}
\label{w1}
\end{figure}

With the branch cut of $\sqrt{u^2 + 1}$ going from $-i$ to $i$ we can take that of the logarithm to run along the positive $u$ axis and deform the integration contour away to infinity as shown in figure \ref{w1}.  The two horizontal segments do not quite cancel because of the discontinuity in the logarithm, but together they yield an easily calculable contribution.  Likewise, the contribution from large $u$ also does not vanish, but is easily calculable.  The integral, in the limit of $r$ going to infinity, is then

\begin{equation}
\begin{split}
-{i \over \pi} \left[ (2 \pi i) \log 2r - 2 \pi^2 - 2 \pi i \int_{u_0}^r {du \over \sqrt{u^2 + e^{2 \pi \eh}}} \right] = \\
2 \pi i + 2 \pi \eh + 2 \log \left( {\sinh \pi k \over e^{\pi \eh}} + \sqrt{1+{\sinh^2 \pi k \over e^{2 \pi \eh}}} \right)
\end{split}
\end{equation}and so

\begin{equation}
h(k+2i) = \left( {\sinh \pi k \over e^{\pi \eh}} + \sqrt{1+{\sinh^2 \pi k \over e^{2 \pi \eh}}} \right)^2 h(k) \label{eq:f(k)}
\end{equation}As one can straightforwardly verify, this equation relating $h$ to its translates by $2i$ is the essential ingredient showing that (\ref{eq:soln}) has the same branching structure as $\sqrt{\sinh^2 k + e^{2 \pi \eh}}$ and so is well defined on the same Riemann surface.  Now it turns out that the function of $k$ obtained by summing (\ref{eq:soln}) over the two preimages of each $k$ in the Riemann surface is precisely the the antisymmetrization discussed at the beginning of this section.  Specifically, the function in question is

\begin{equation}
u(k) = h_{+}(k) + h_{-}(k)
\end{equation}with
\begin{equation}
\begin{split}
h_{+}(k) = {\sinh {\pi k} \over \sqrt{\sinh^2 {\pi k} + e^{2 \pi \eh}}} \exp \left\{ \pi i \int_{k_0}^k {dk' \sinh {\pi k'} \over \sqrt{\sinh^2 {\pi k'} + e^{2 \pi \eh}}} \left(\eh - {k' \over \tanh \pi k'} \right) \right\}\\
h_{-}(k) = - {\sinh {\pi k} \over \sqrt{\sinh^2 {\pi k} + e^{2 \pi \eh}}} \exp \left\{ - \pi i \int_{k_0}^k {dk' \sinh {\pi k'} \over \sqrt{\sinh^2 {\pi k'} + e^{2 \pi \eh}}} \left(\eh - {k' \over \tanh \pi k'} \right) \right\} \label{eq:fullsoln}
\end{split}
\end{equation}where $k_0 = {i \over \pi} \sin^{-1} \left(e^{\pi \eh} \right)$.

From its construction as a sum over preimages we immediately see that $u(k)$ is analytic everywhere (the endpoints of the cuts are the only places where one must take a little care).  The fact that it solves (\ref{eq:eigen}) will be checked presently.

\section{Checking the Solution}
We explicitly verify that $u(k)$ is a solution.  We multiply (\ref{eq:eigen}) through by $\left( 1 - e^{-2i \partial_k} \right)$ to obtain

\begin{equation}
\left( 1 - e^{-2i \partial_k} \right) \left( {k \over \tanh \pi k} - \eh \right) u(k) = {1 \over \pi} \left(1 + e^{-2i \partial_k} \right) i \partial_k u(k) \label{eq:tocheck}
\end{equation}Now, the operator $e^{-2i \partial_k}$ is just translation by $-2i$ when it acts on $u(k)$.  When it acts on $h_{\pm}(k)$ it is not well defined because the branch cuts make the expansion in derivatives fail to converge.  Nevertheless, we will prove the above equation is true for $h_{\pm}(k)$, where we will interpret $e^{-2i \partial_k}$ to mean translation by $-2i$.  This will show that the above is true for $u(k)$, because in that case $e^{-2i \partial_k}$ is both well defined and translation by $-2i$.  Let's take $h_{+}(k)$ for definiteness.

Recall that $x = {e^{2\pi \eh} / (e^{\pi k} - e^{-\pi k})^2}$.  We have

\begin{equation}
{i \over \pi} \partial_k h_{+}(k) = -{1 \over \sqrt{1+4x}} \left( \eh - {k \over \tanh \pi k} \right) h_{+}(k) + {4xi \over 1+4x} {1 \over \tanh \pi k} h_{+}(k)
\end{equation}Recall also (\ref{eq:f(k)}) that $e^{-2i\partial_k}$ acting on $h_{+}(k)$ multiplies it by

\begin{equation}
f = {1\over 2x} \left( 1+2x-\sqrt{1+4x} \right)
\end{equation}Given these two facts, we compute that the right and left hand sides of (\ref{eq:tocheck}) (with $u(k)$ replaced by $h_{+}(k)$) are, respectively,

\begin{equation}
\begin{split}
-\left( 1 \over \sqrt{1+4x} \right) \left( \eh - {k \over \tanh \pi k} \right) \left(1+f\right) h_{+}(k) - {2if \over \sqrt {1+4x}}{1 \over \tanh \pi k} h_{+}(k) + \\
{4xi \over 1+4x}{1 \over \tanh \pi k} \left(1+f\right) h_{+}(k)
\end{split}
\end{equation}and

\begin{equation}
(f-1) \left(\eh - k \over \tanh \pi k \right) h_{+}(k) + {2if\over \tanh \pi k} h_{+}(k).
\end{equation}Inserting the expression for $f$ and using elementary algebra shows that these are indeed equal.

\section{The phase shift}

First, we examine the large $k$ behavior of the exponent of, say, $h_{+}$.  We denote it by $F(k,\eh)$:

\begin{equation}
F(k,\eh) = \pi i \int_{k_0 \left(\eh \right)}^k {dk' \sinh {\pi k'} \over \sqrt{\sinh^2 {\pi k'} + e^{2 \pi \eh}}} \left(\eh - {k' \over \tanh \pi k'} \right).
\end{equation}where as before $k_0 \left(\eh \right) = {i \over \pi} \sin^{-1} \left(e^{\pi \eh} \right)$.  This integral cannot be done in closed form, but by differentiating with respect to $\eh$ under the integral sign one does obtain one that can be done through simple integration by parts.  The result is

\begin{equation}
\partial_{\eh} F(k,\eh) = \left[ {\pi i \sinh \pi k \over \sqrt{ \sinh^2 {\pi k} + e^{2 \pi \eh}}} \left( k + {\eh e^{2 \pi \eh} \over 1-e^{2 \pi \eh}} {1 \over \tanh {\pi k}} \right) \right]_{k_0 \left(\eh \right)}^k.
\end{equation}This can be evaluated quite easily if one notices that it is simply half of the integral from $-k$ to $k$, with the branch cut chosen to go from $-k_0(\eh)$ to $k_0(\eh)$ and the contour deformed above it.  Thus

\begin{equation}
\partial_{\eh} F(k,\eh) = {\pi i \sinh \pi k \over \sqrt{ \sinh^2 {\pi k} + e^{2 \pi \eh}}} \left( k + {\eh e^{2 \pi \eh} \over 1-e^{2 \pi \eh}} {1 \over \tanh {\pi k}} \right).
\end{equation}Taking the limit of $k$ going to infinity we see that $F(k,\eh)$ approaches $\pi i \left(k + {\eh e^{2 \pi \eh} / \left(1- e^{2 \pi \eh}\right)} \right)$, which is starting to look like the conjectured phase shift (\ref{eq:phase}).  

Let us now figure out some asymptotics.  We spare the reader the easy integrals and state only that in the limit $\eh \ll 0$ we have

\begin{equation}
F(k, \eh) \cong \pi i \left(\eh k - {1 \over 2} k^2 - {1 \over 12} \right)
\end{equation}and in the limit $\eh \gg 0$

\begin{equation}
F(k, \eh) \cong \pi i \left( -{1 \over 2} (k-\eh)^2 - {1 \over 6}\right).
\end{equation}Because we know the derivative of $F(k, \eh)$ with respect to $\eh$, we can figure out the phase shift for any energy $\eh$.  We obtain that for large $k$,

\begin{equation}
u(k) \cong \sin \left[ \pi \left( \eh k - {1\over 2} k^2 - {1\over 2} \int_{-\infty}^{\eh} d\epsilon' \left( \epsilon' + {\epsilon' \over \tanh \pi \epsilon'} \right) \right) \right]
\end{equation}This matches (\ref{eq:phase}).

%First, we note that (\ref{eq:eigen}) is self-dual under fourier transforms.  Specifically, the $h$ that satisfies (\ref{eq:eigen}) also satisfies
 
%\begin{equation}
%\left( {\pi \left({\tau \over \pi}\right) \over \tanh \left({\tau \over \pi}\right)} - {i \partial_{\left({\tau \over \pi}\right)} \over \tanh i \partial_{\left({\tau \over \pi}\right)}} \right) h = - \pi \eh h
%\end{equation}where $\tau=i \partial_k$.

%%%%%%%%%%%%%%%%%%%%%%%%%%%%%%%%%%%%%%%%%%%%%%%%%%%%%%%%%%%%%%%%%%%

\section*{Acknowledgments}

I would like to thank A.~Frolov, J.~Maldacena, A.~Maloney, J.~McGreevy, and S. Shenker for useful conversations.  This work was supported by NSF grant 0244728.
%\bibliographystyle{my-phys}
%\bibliography{string}

\end{document}